\documentclass[apjl]{emulateapj}
\usepackage{psfig,amsfonts,amsmath,graphicx,natbib}
\def\eiso{\ifmmode {E_{iso}}\else
                ${E_{iso}}$\fi}
\def\pc{\ifmmode {\rm pc}\else
                pc\fi}
\def\sfr{\ifmmode \dot{{\rm M}}_\star \else
                $\dot{{\rm M}}_\star$\fi}
\def\mstar{\ifmmode {\rm M}_\star \else
                ${\rm M}_\star$\fi}
\def\msun{M_\odot}

\def\erg{{\rm erg}}
\def\cm{{\rm cm}}
\def\De{{\cal D}_{est}}

\begin{document}

\title{Linking Short Gamma Ray Bursts and their Host Galaxies}

\author{
James E. Rhoads\altaffilmark{1}
}

\altaffiltext{1}{Arizona State University}
\begin{abstract} 
The luminosities of SGRB host galaxies are anticorrelated with both
the isotropic equivalent gamma ray energy and the gamma ray luminosity
of the explosions.  Observational selection effects only strengthen
the significance of this correlation.  The correlation may indicate
that there are two physically distinct groups of SGRBs.  If so, it
requires that the more luminous class of explosions be associated with
the younger class of progenitors.  Alternatively, it could be due to a
continuous distribution of burst and host properties.  As one possible
explanation, we find that the effect of binary neutron star masses on
inspiral time and energy reservoir produces a correlation of the
appropriate sign, but does not automatically reproduce the correlation
slope or the full range of SGRB energy scales.  Any future model of
SGRB progenitors needs to reproduce this correlation.
\end{abstract}


\section{Introduction}
\label{sec:intro}
The host galaxies of short gamma ray bursts (SGRBs) 
exhibit a wide range of physical properties.
SGRBs have been found not only in star-forming
hosts, but also in elliptical galaxies with strong
upper limits on their star formation rate (hereafter ``quiescent
hosts'') (Prochaska et al 2006; Berger 2008).  
In contrast, long gamma ray bursts
(LGRBs) are found exclusively in star forming galaxies
(e.g., Le Floc'h et al 2003, Fruchter et al 2008, Savaglio et al 2009).
Several lines of evidence now link LGRBs
with the deaths of massive stars (Hjorth et al 2003,
Stanek et al 2003, Fruchter et al 2006, Raskin et al 2008).  
The origin of the SGRBs
is not firmly established, but suspicion centers on merging neutron
star binaries, or other phenomena primarily associated
with neutron stars (Belczynski et al 2006; Chapman, Priddey, 
\& Tanvir 2008).  Hosts offer one of the best tools for unveiling
the nature of the GRB progenitors, because the properties of the gamma
ray emission and afterglow are decoupled from many important details
of the progenitor objects under the fireball model of GRBs (e.g.
Paczy\'{n}ski \& Rhoads 1993, Katz 1994, Piran 2005).

The situation is similar to supernovae, which divide into two
groups, the type Ia and the core collapse events (which
include types II, Ib, and Ic).
Core collapse events are found only in star forming galaxies, 
and are associated with the deaths of stars having initial
masses $\ga 8 \msun$. In contrast, type Ias are found in both quiescent 
and star forming hosts, and are thought to be powered by the nuclear 
explosion of a white dwarf whose mass is pushed above the Chandrasekhar
limit by accretion.  However, despite having a characteristic progenitor
mass set by fundamental physics, the Ia supernovae are not entirely 
homogeneous in their properties.  Those found in star forming 
hosts are both more frequent (per unit stellar mass) and more luminous 
than their cousins in quiescent galaxies (Hamuy et al 1995; 
Scannapieco \& Bildsten 2005; Sullivan et al 2006).  

In this {\it Letter}, we look for similar trends linking the 
propertie of SGRBs and their hosts.  We find an anticorrelation 
between SGRB isotropic energy and host galaxy luminosity.  We find 
no obvious selection effects that
could produce such an effect, nor do we find evidence for any similar
effect in a larger control sample of LGRBs.


\section{Observations}
We take our SGRB sample from the compilation by Berger (2008;
hereafter B08) of all SGRBs localized by the X-ray telescope (XRT)
on the Swift satellite (Gehrels et al 2004).  
This sample contains 23 SGRBs, of which 12
have reported redshifts.  These 12 form the core sample for our
study.  They are further divided into 6 with optical
afterglows, and hence the most accurate positions and most secure host
galaxy identifications (``sample 1'' of B08), and 6 with somewhat less
accurate positions based X-ray data alone (``sample 2'' of B08). 
The two subsamples are similar in most properties (B08).

We also form a comparison sample of 34 long bursts, by
combining host galaxy information from Savaglio et al (2008) with
burst durations from the GCN circulars and other GRB properties 
from Amati et al (2008).  

We calculate host galaxy absolute magnitudes as
$B_{abs} = R - 5 \log(d_L / 10 \pc) + 2.5 \log(1+z)$.  Here
R is the observed R-band magnitude from B08, and
$d_L$ is the luminosity distance,
calculated assuming $H_0 = 71 \hbox{km} \hbox{s}^{-1} 
\hbox{Mpc}^{-1}$,  $\Omega_{0,m} = 0.27$, and $\Omega_{0,\Lambda} =
0.73$.  The resulting absolute magnitude corresponds to
rest wavelength $6500 \hbox{\AA} / (1+z)$.  For 
our SGRB sample, this always between U and R band, and typically
near B band (hence our notation).
We calculate the isotropic equivalent gamma ray energy
of the bursts directly from their fluence $f$ as
$\eiso = 4\pi d_L^2 f / (1+z)$.

\section{Results}
\begin{figure}
\centering
\includegraphics[angle=0,scale=0.49]{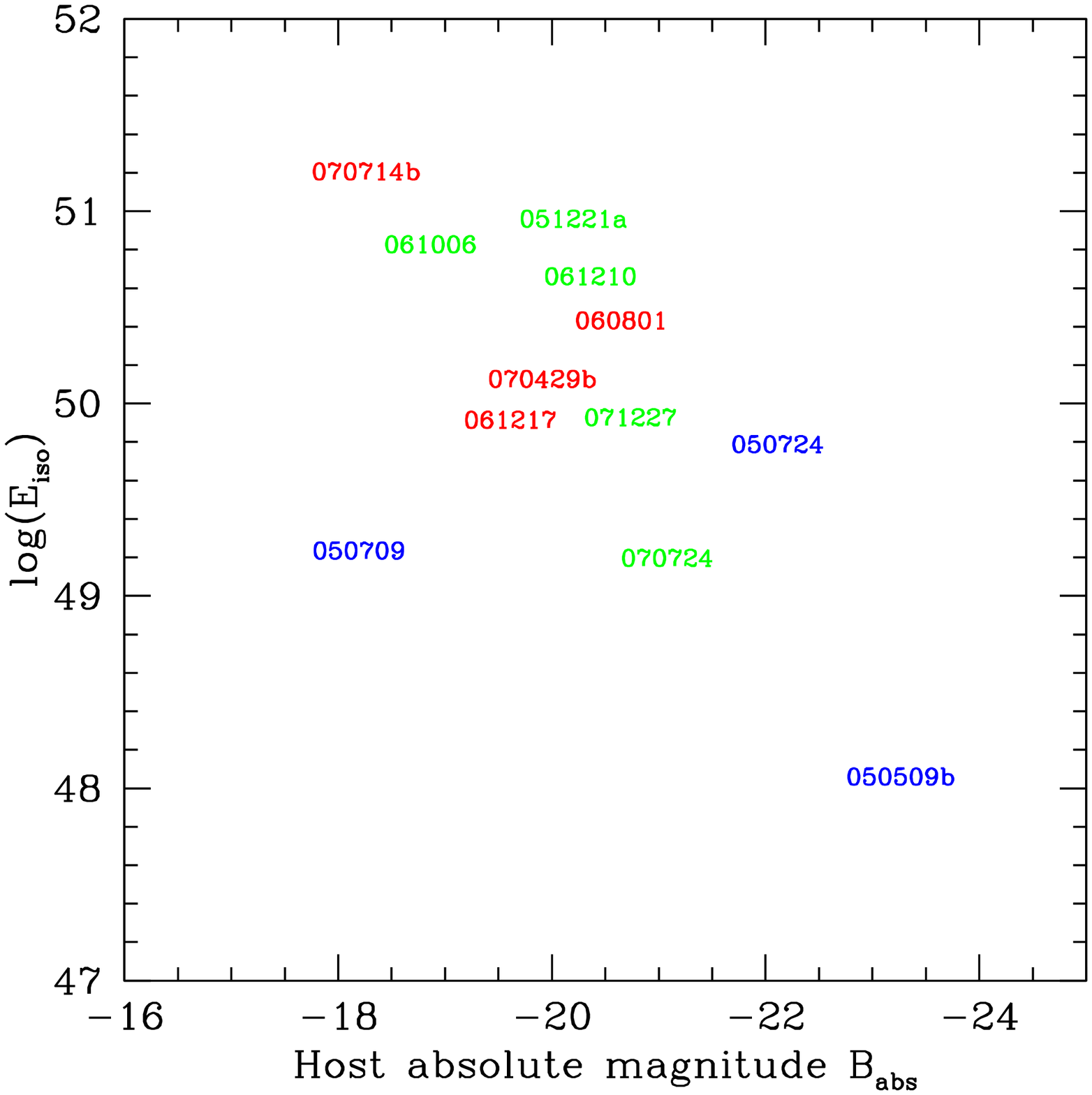}
\includegraphics[angle=0,scale=0.49]{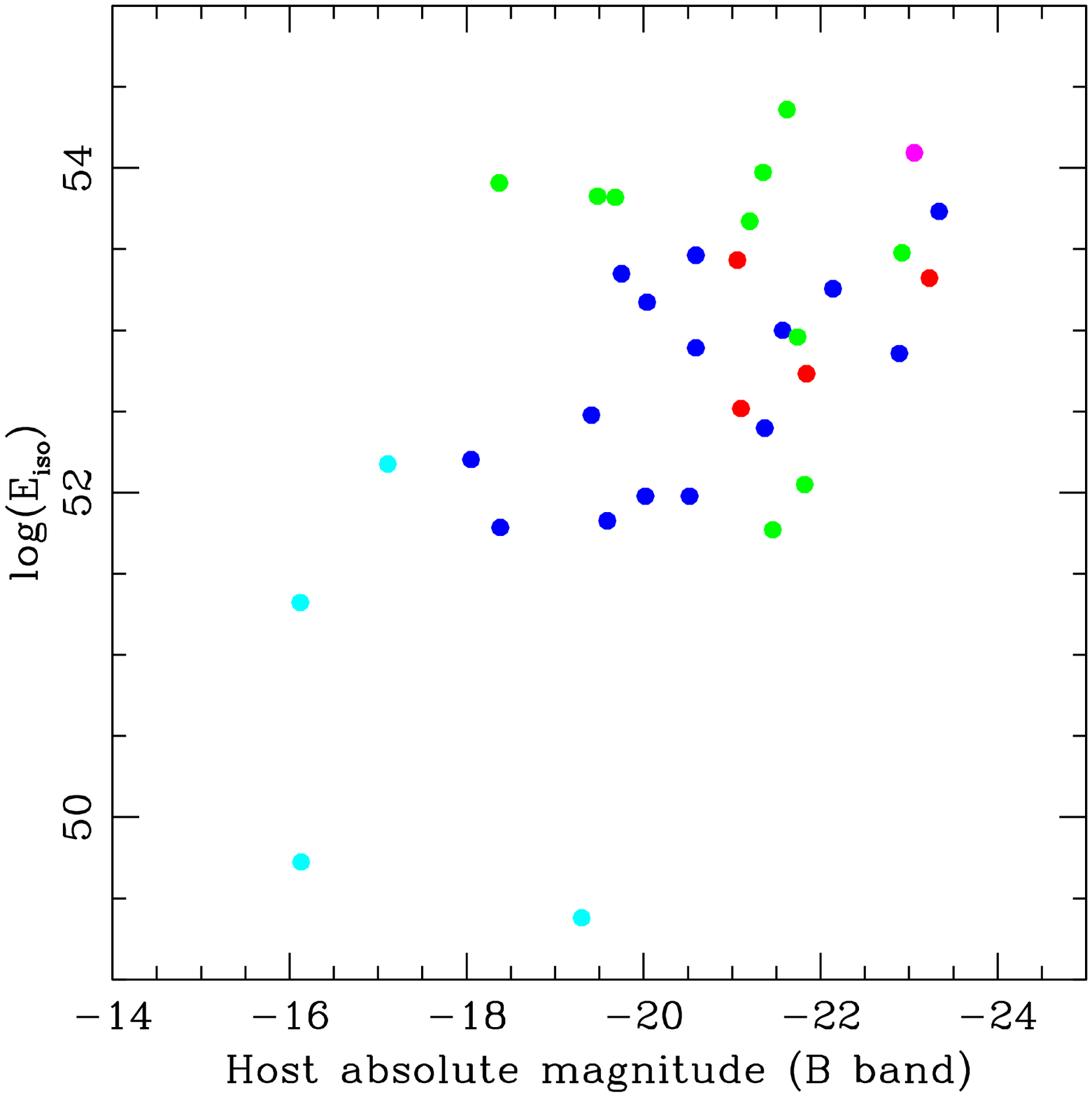}
\caption{
The relation between isotropic equivalent gamma-ray energy and host
galaxy absolute magnitude for short GRBs (first panel) and long GRBs
(second panel).  The
observed correlation in the SGRB sample is significant at about
the 99\% level (see text).
The control sample of LGRBs shows a correlation
consistent with that expected from observational selection.  SGRB data
points are labelled with the GRB ID number.  Redshift ranges are denoted
by color.  For SGRBs, blue means $z<0.3$, green
means $0.3<z<0.7$, and red means $z>0.7$.  For LGRBs, cyan means $z<0.3$,
blue means $0.3<z<1.0$, green means $1.0<z<2.0$, red
means $2.0<z<3.5$, and magenta means $z>3.5$.
\label{fig:eiso_absmag}}
\end{figure}

The isotropic-equivalent energy of the SGRBs is correlated
with the absolute magnitudes of their host galaxies, with a
correlation coefficient of 0.60.  That is, 
the brightest bursts tend to occupy the least luminous hosts, and vice
versa (fig.~\ref{fig:eiso_absmag}a).  This is {\it not} expected from
simple observational selection effects.  If the samples are largely
limited by flux and/or fluence, with most objects near the
detection threshold, the range of distances in the sample should then
lead to a {\it positive} correlation of burst energy with host
luminosity.  Such a positive correlation {\it is} seen for our
control sample of long GRBs (figure~\ref{fig:eiso_absmag}b).

The least luminous SGRBs are the closest, while the 
least luminous hosts span a wide range of
redshifts. Gamma ray sensitivity limits inclusion in our sample
more critically than does sensitivity to the host galaxy's starlight.
If the host is not detected in the first relevant observation,
more data can be taken, but if the GRB is not detected in the
first observation, it can never enter the sample.

If a low-luminosity event such as GRB 050509B occurred in
a subluminous host at $z \la 0.5$, we should still see
it.  We see perhaps one such event, GRB 070724, which has
low isotropic energy and luminosity, but lives in a star
forming host galaxy. 
We would also expect that if a highly luminous SGRB occurred
in a large elliptical host galaxy, we should be able
to observe such an event (and its luminous host) out to $z\ga 0.8$.  
We see no such events.  This suggests that
the more luminous SGRBs are associated with young progenitors
of some type that is absent in old elliptical galaxies.

We have tested the significance of the correlation in 
figure~\ref{fig:eiso_absmag}a using a Spearman rank correlation
test.  The test yields a rank correlation coefficient of $0.50$
and a significance level of $t=1.84$ with 10 degrees of
freedom.  Formally, this corresponds to a 95\% significance.   

However, the true significance is higher, since observational
selection effects work to produce the opposite sign of
correlation.  We have performed simple
simulations of the correlation produced by
observational selection in the absence of any true, physical
relation between the burst and host properties.
We first constructed independent probability distributions for
$\eiso$, $B_{abs}$, and $z$.  To these, 
we added selection probabilities based on gamma-ray 
fluence and host apparent magnitude.  Each of
these probability distributions was specified by a simple,
plausible functional form.  The functions were adjusted until
the simulations were able to reproduce separately the 1D marginal
distributions of $f$, $\eiso$, $R$, $B_{abs}$, and $z$
seen in the data (all within the Poisson uncertainties).

In each simulation, we constructed a ``parent
sample'' of bursts, and applied the 
selection probabilities to simulate a ``selected sample,'' whose
size is constrained to equal that of the observed sample.
We then measure the correlation coefficient between \eiso\ 
and $B_{abs}$ in the simulation.  We repeat this procedure 
many times, and compare the distribution of simulated correlation 
coefficients to that measured in the observed
sample. 

Among $10^4$ simulations, only 121 have coefficients above the observed
value.  Thus, by considering selection effects, we estimate the
significance of the SGRB energy - host luminosity anticorrelation
at about 98.8\%. 

While the simulations include several {\it ad hoc} functional
forms, they capture the essential points of any realisitic model
that does {\it not} have correlations between burst and host
properties.  First, the bursts and hosts span a wide range
of intrinsic brightness.  Second, the rate as a function of
redshift could be matched by the volume element modified by some
modest amount of rate evolution.  Third, the observational selection
functions are monotonically decreasing over the range from the
brightest to the faintest observed SGRBs and hosts.
Fourth, the selection on the SGRB fluence excludes many more simulated
events than does the selection on host luminosity.  This induces
a relatively weak correlation between host and burst brightness
in the simulated samples.  More strongly peaked intrinsic properties,
or a more even balance between rejection by either GRB or host
properties, would tend to induce stronger correlations in the
simulated sample.  Thus, plausible changes
in our simulations should not strongly affect our
conclusions.

Our results do depend substantially on GRB 050509b: If we
exclude it from the sample, the remaining 11 observed
points yield a correlation coefficient of 0.28.  This corresponds to
about a 14\% chance in the simulations. 
For comparison, a hypothetical sample of about 40 SGRBs with a
true correlation coefficient 0.28 would be significant at 
the same the 99\% level as our full sample.
Provided that the identification of the GRB 050509b host is
regarded as secure, the correlation does not depend
critically on other bursts without optical transients. If we use
the subset of SGRBs with optical transients (B08's ``sample 2'')
plus GRB 050509b (which has no OT but still a rather secure
host detection), the correlation coefficient becomes 0.63,
and the simulations show a significance level of 95.6\%.

For reference, our control sample of LGRBs shows a correlation
coefficient of -0.54 between isotropic energy and host absolute
magnitude.  We adapted our simulations to reproduce the marginal
distributions of $z$, $R$, $B_{abs}$, and so forth for the LGRB
sample, and again compared the distribution of correlation
coefficients to that observed for the LGRBs.  Formally, this too
yields a difference at the 99\% level between the simulations and the
data, here in the sense that the data show a stronger {\it positive}
correlation between GRB and host brightness than do the simulations.
Because the sign of the correlation matches the sign of the selection
effects, and because the correlation is driven substantially by the
X-ray flashes (020903 and 060218) at the bottom of the LGRB \eiso\ 
distribution, we do not attach great physical significance to this
result yet.  However, it is intriguing and may merit further
investigation in future.

We also examined the relation between SGRB gamma ray luminosity
and host galaxy star formation
rate per unit stellar mass.  Because the two most luminous SGRB hosts
are early type galaxies, with tight upper limits on their emission-line
derived SFR (Prochaska et al 2006, Berger 2008,
Savaglio et al 2008), this plot clearly separates these two 
SGRBs from the rest of the sample.  The specific star formation
rate is defined as $s = \sfr / \mstar$, where $\sfr$ is the star
formation rate, and $\mstar$ the stellar mass of the galaxy.  However,
lacking the data to fit an accurate stellar mass to some galaxies,
we use the quantity $\log({\cal S}) \equiv \log{\sfr} + 0.4
B_{abs} = \log{(s)} + \log{(\mstar / L_{host})}$.  To
demonstrate the utility of ${\cal S}$, we compared it to the
published $s$ (from Savaglio et al 2008) for the 34 LGRB 
hosts from our control sample.
We find $\log({\cal S}) \approx 0.4 \log(s \times \hbox{Gyr}) - 7.9$,
with a scatter of $0.3$ dex (RMS).

\begin{figure}
\includegraphics[angle=0,scale=0.49]{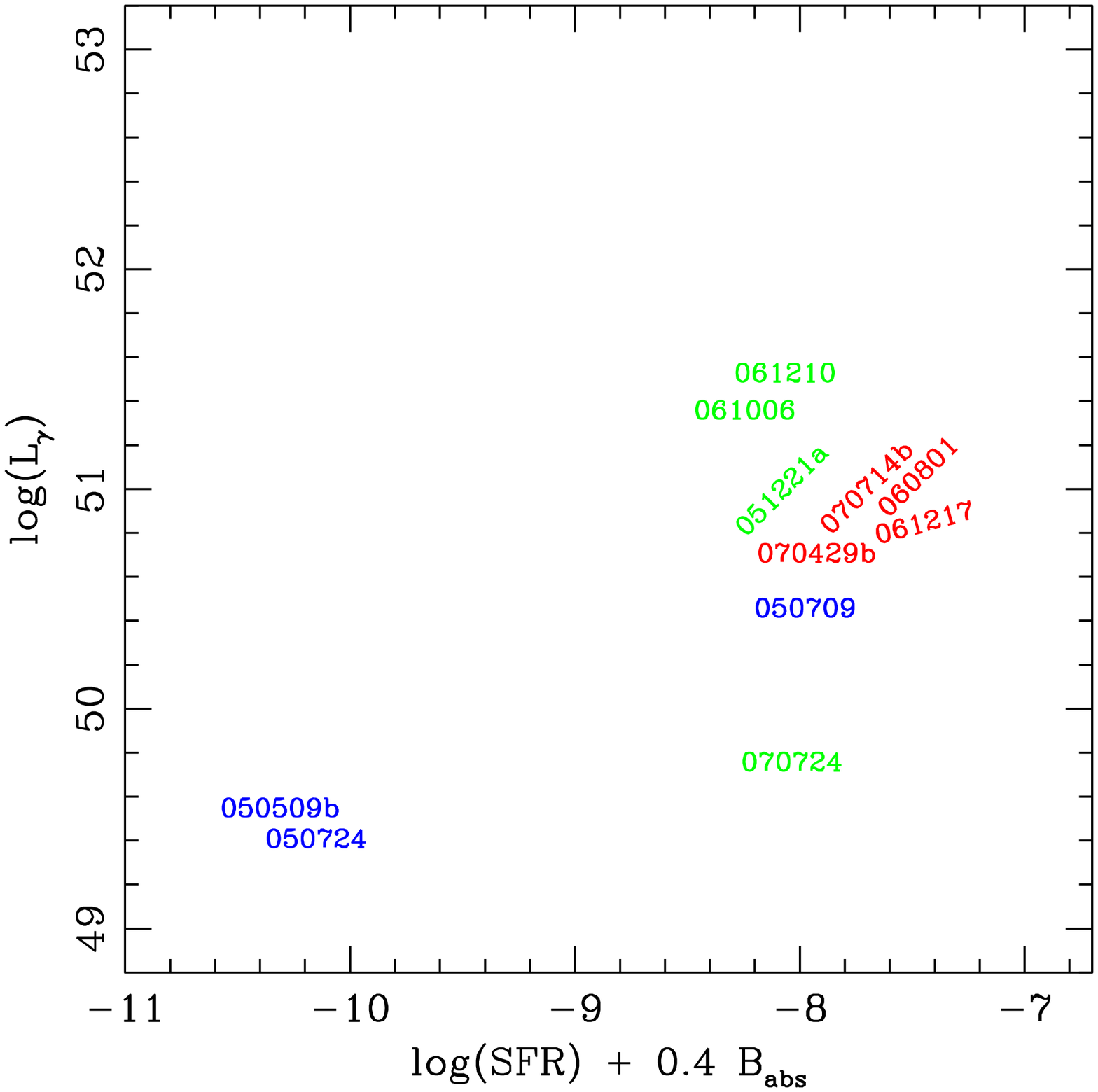}
\includegraphics[angle=0,scale=0.49]{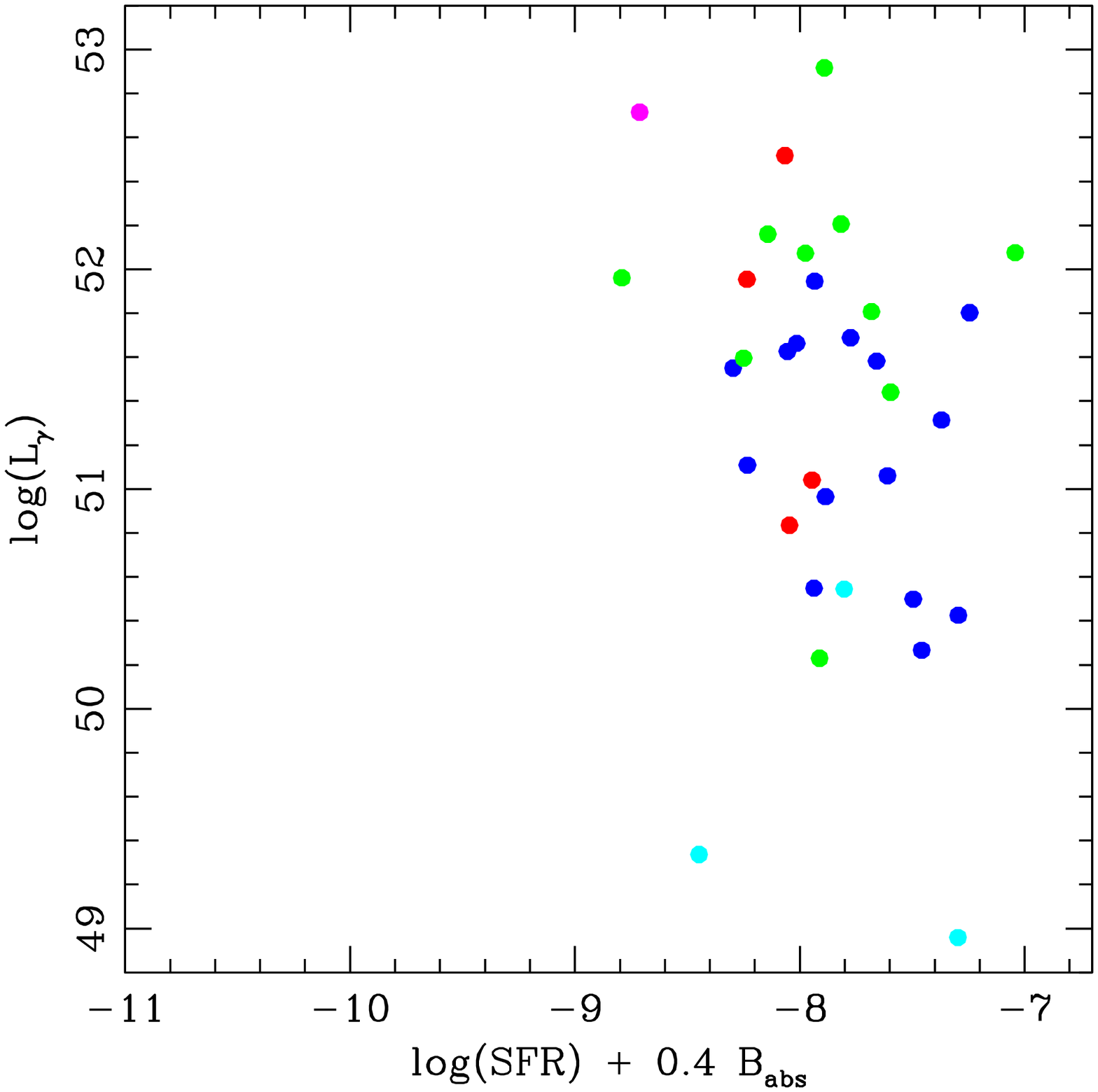}
\caption{The relation between GRB luminosity
and the specific star formation rate of the host galaxy, for
short (first panel) and long (second panel) GRBs.  
The least luminous SGRBs are the two in massive early-type 
host galaxies. While some bursts in star forming hosts 
appear comparably faint (i.e., GRB 070724), we have yet to 
observe a bright burst in a quiescent host.
For the long GRBs, there is no visible correlation of
$L_\gamma$ and specific star formation rate.
Point labels follow the same definitions as in figure~1.
\label{fig:ssfr_lgam}}
\end{figure}

The results (fig.~\ref{fig:ssfr_lgam}a) show that the
separation of the SGRB hosts into actively star forming galaxies
and old stellar populations is accompanied by a
separation of burst gamma-ray luminosity, with only the least luminous
bursts found in the old hosts.  The corresponding plot for 
LGRBs (fig.~\ref{fig:ssfr_lgam}b) shows no correlations beyond 
the sample selection requirement that the least luminous events 
be reasonably nearby.  A comparison of the two figures also shows
the tendency (noted previously by, e.g., B08) for even star-forming
SGRB hosts to have less vigorous star formation (i.e. lower ${\cal S}$)
than do LGRB hosts, and also the tendency for SGRBs to have lower
gamma ray fluxes.

\section{Discussion}
While our sample is small, the 
anticorrelation between SGRB $\eiso$ 
and host luminosity is probably real, based on our Monte Carlo
simulations.  The similar anticorrelation between brightness of 
type Ia supernovae and
the stellar population ages of their host galaxies was first described
on the basis of a similarly small sample (13 supernovae; Hamuy et al
1995). 
We now examine possible interpretations for this
result.

Suppose, first, that two distinct mechanisms produce
SGRBs.  Such a situation would arise
naturally if two or more of the different proposed progenitors
actually do produce SGRBs.  Combinations that have been explored 
include coalescence of double neutron star binaries and
of NS-black hole binaries (Belczynski et al 2006; O'Shaughnessy et al 2008;
Troja et al 2008), and combinations of NS-NS binaries
with giant flares from magnetars (Chapman, Priddey, \& Tanvir 2008).  
Additionally, a range of SGRB
properties might arise from NS-NS binaries alone, provided those
binaries evolved through a range of different binary interaction
histories (Belczynski et al 2006; Salvaterra et al 2008).

If one mechanism is associated with stars younger than 1 Gyr, 
and the other with ancient populations (age $\sim H_0^{-1}$), 
we can explain the properties of figures~\ref{fig:eiso_absmag}a
and \ref{fig:ssfr_lgam}a. 
The fraction of old-progenitor SGRBs occurring in
quiescent galaxies should match the fraction of stellar mass
in quiescent galaxies.  The present, 
limited data shows two bursts in
quiescent hosts, and a third similarly low-luminosity burst (070724)
in a star-forming host galaxy. This matches well the fraction of
stellar mass in elliptical galaxies at low redshift, which is 54\% to
60\% (Baldry et al 2004).
Conversely, the fraction of more luminous SGRBs occurring
in quiescent hosts is $0/9$ in the present sample 
(fig.~\ref{fig:ssfr_lgam}a).  Under a two-population scenario,
the mechanism responsible for the more luminous SGRBs
should have a negligible rate for ages exceeding a few Gyr.

If instead there is a single, continuous distribution of SGRB 
and host properties, it suggests that the SGRB progenitor's
properties vary systematically with either age or metallicity,
which vary systematically along the Hubble sequence and 
have direct effects on stellar-scale physics.

If SGRBs are due to binary neutron star
inspiral events, an anticorrelation of SGRB
energy and the specific star formation rate of the host
galaxy could come from the dependence of inspiral time and
available energy on mass. Gravitational radiation gives
an inspiral time of
\begin{equation} 
t_{inspiral} = {5 c^5 \over 256 G^3} {a_0^4 
\over M_1 M_2 (M_1 + M_2)} ~
\end{equation}
(e.g., Landau \& Lifschitz 1958), where
$M_1$ and $M_2$ are the masses of the two bodies,
$a$ is the semimajor axis of their orbit, and $G$ and $c$ are Newton's
constant and the speed of light.  
Thus, if a star formation event yields a population of compact object 
binaries with a range of mass, we expect a range of inspiral times.
The SGRBs occurring in quiescent hosts would be those with low masses and/or 
wide initial separations.  The available energy reservoir is 
$E \sim G M_1 M_2 / R_{final}$, where $R_{final}$
is the characteristic size of the system when the GRB energy 
is released.  If the SGRB event produces a black hole,
we expect $R_{final} \sim R_{grav} \propto (M_1+M_2)$,
so that $E\sim M_1 M_2 c^2 / (M_1 + M_2)$.  Then, for $M_1 \approx M_2$,
$t_{inspiral} \sim a_0^4 E^{-3}$.   
Characteristic stellar ages for SGRB hosts range
from $\sim 10$ Gyr for quiescent hosts, to
$\sim s^{-1} \sim 1$ Gyr for typical star forming SGRB hosts
(see, e.g., Savaglio et al 2008).
If this tenfold range of age corresponds to a tenfold range
of $t_{inspiral}$, it would imply a range of $\sim 2\times$ in 
neutron star mass, which is comparable to the range of NS masses 
believed to exist in our Galaxy.

However, the corresponding $2\times$ range of energy 
cannot explain the range of \eiso\ observed in 
figure~\ref{fig:eiso_absmag}a.  A larger range in SGRB energetics 
might follow if $R_{final} \sim R_{NS}$, especially for a
neutron star equation of state that gives $d R_{ns} / dM < 0$.
Moreover, $a_0$ may correlate with $M_1$ and $M_2$ in some
complex way depending on on both binary star mass transfer and 
detailed supernova physics.  While a more thorough understanding
of the SGRB mechanism is clearly required before we can fully
model the relation between host and SGRB properties, the existence
of such a relation will provide valuable clues to the nature of
the short GRBs.

The correlation in figure~\ref{fig:eiso_absmag}a can also 
provide a tool for estimating SGRB redshifts.
$\log(\eiso)$ and $B_{abs}$ 
are related empirically by 
$\log(\eiso / \erg) \approx 50 + 0.725 (B_{abs} + 20.2)$.
Defining ${\cal D} = d_L / \sqrt{1+z}$, so that
$\eiso = (4\pi {\cal D}^2) f$ and 
$B_{abs} = R_{AB} - 5 \log[({\cal D} / 10pc)]$,
we can substitute for $\eiso$ and $B_{abs}$ 
to obtain a distance estimate:
\begin{equation}
\log(\De / \hbox{Gpc}) = 
0.012 + 0.129 (R_{AB}-20) - 0.177 \log(f_{-6}) ~~,
\label{eq:dist_est}
\end{equation}
where $f_{-6}$ is the SGRB fluence in units of $10^{-6} \erg \, \cm^{-2}$.
The conversion from $\De$ to the estimated distance $d_{L,est}$ and redshift
$z_{est}$  must of course follow the cosmology we used to derive 
equation~\ref{eq:dist_est}.
For our primary sample of 12 SGRBs, the rms residual in
$\log(\De)$ is $0.17$ dex, and in
redshift $z(\De)$ is 0.147.  
The residuals from the fit substantially exceed the uncertainties
expected from fluence and host magnitude measurements, implying that
most of this scatter is intrinsic.

As an alternative, we can directly fit a relation of the form
$\log(z_{est}) = \log(z_0) + a (R_{AB} - 20) + b \log(f_{-6})$
by minimizing $\Delta z_{rms} = 
\langle (z_{est} - z_{obs})^2 / (N-1)\rangle^{1/2}$.
This approach gives $\log(z_{est}) = -0.56 + 0.107 (R_{AB}-20) - 0.14  
\log(f_{-6})$, with $\Delta z_{rms} =  0.142$. 
This redshift estimator is easier to apply, and is formally independent
of the adopted cosmological parameters, though it is less directly linked
to whatever physical mechanism links $\eiso$ and $B_{abs}$.  Differences
between the predicted redshifts from the two equations are all much
smaller than the scatter in either.

Another possibility is that the external medium around
SGRBs plays some role in their luminosity.  The interstellar medium is
typically less dense in elliptical galaxies than in actively star
forming galaxies.  However, this should only affect the
observed flux from external shocks that GRB ejecta drive into the
ambient medium.  External shocks are believed to power the afterglow
emission, but not the GRB luminosity itself (e.g., Piran 2005).  Since
our observed anticorrelation involves the SGRB $\gamma$-ray flux itself,
we disfavor this explanation.

To conclude, we have found an anticorrelation between the energy of a
short gamma ray burst and the luminosity of its host galaxy.  Such an
anticorrelation occurs with a probability of $\sim 1\%$ in simulations that
account for observational selection effects, and 
so is probably real.  Its physical origin 
is unclear, though it is most likely due to a correlation between the
age of an SGRB progenitor and the luminosity of the explosion.  If
this correlation is a continuous distribution, it can provide
approximate redshift estimates for SGRBs.  Time will tell 
whether this correlation holds for a larger sample, 
and whether it is a property of a single
distribution, or if it instead reflects an underlying division of
SGRBs into two physically distinct sets.

\acknowledgements
I thank Sangeeta Malhotra, Evan Scannapieco, Patrick Young, Sumner
Starrfield, and Frank Timmes for stimulating discussions; 
Sandra Savaglio and Edo Berger for their published host galaxy
compilations; R. Quimby, E. McMahon, and J. Murphy
for the GRBlog database at
http://grad40.as.utexas.edu/grblog.php ; and the Max
Planck Institute for Astronomy for hospitality
during completion of this work.


\end{document}